\documentclass[amsfonts,showpacs,tightenlines,aps,12pt,floatfix]{revtex4}
\usepackage{bm}
\usepackage{epsfig}
\begin{document}


\title{STATUS OF THE HADRONIC $\tau$ DETERMINATION OF $\vert V_{us}\vert$}

\author{K. Maltman\footnote{alternate address: CSSM, School of Chemistry
and Physics, University of Adelaide, SA 5005 Australia}}

\affiliation{Department of Mathematics and Statistics, 
York University, 4700 Keele St.,\\
Toronto, Ontario M3J 1P3, Canada}

\author{C.E. Wolfe}

\affiliation{Department of Physics and Astronomy, 
York University, 4700 Keele St.,\\
Toronto, Ontario M3J 1P3, Canada}

\author{S. Banerjee, J.M. Roney and I. Nugent}

\affiliation{Department of Physics and Astronomy, University of Victoria,\\
PO Box 3055, STN CSC, Victoria, BC V8W 3P6, Canada}

\date{\today}

\begin{abstract}
We update the extraction of $\vert V_{us}\vert$ from hadronic $\tau$ decay
data in light of recent BaBar and Belle results on 
the branching fractions of a number of important strange decay modes. 
A range of sum rule analyses is performed, particular attention being 
paid to those based on ``non-spectral weights'', developed previously 
to bring the slow convergence of the relevant integrated $D=2$ OPE series 
under improved control.  Results from the various sum rules are shown
to be in good agreement with one another, but $\sim 3\sigma$ below 
expectations based on 3-family unitarity. Important avenues for future 
study are briefly discussed.
\end{abstract}


\maketitle

\section{Introduction}	
Once B factory data is fully analyzed, the precision 
of the hadronic $\tau$ decay determination of $\vert V_{us}\vert$
is expected to match 
or exceed that of $K_{\ell 3}$- and 
$\Gamma [K_{\mu 2}]/\Gamma [\pi_{\mu 2}]$-based methods~\cite{gamizetal,kmcw},
providing important tests of three-family unitarity
and the Standard Model (SM) scenario for weak semi-leptonic $\Delta S=1$ 
interactions. The analysis involves weighted integrals of the spectral 
functions of the vector (V) and/or axial vector (A) current-current 
correlators, examples of which are the normalized branching function ratios 
$R_{V/A;ij}\, \equiv\, \Gamma [\tau^- \rightarrow \nu_\tau
\, {\rm hadrons}_{V/A;ij}\, (\gamma )]/ \Gamma [\tau^- \rightarrow
\nu_\tau e^- {\bar \nu}_e (\gamma)]$ associated with inclusive flavor 
$ij=ud,us$, V/A-current-induced hadronic $\tau$ decays.

The ratios $R_{V/A;ij}$ are related to the $J=0,1$ spectral functions,
$\rho_{V/A;ij}^{(J)}(s)$, of the corresponding current-current 
correlators via~\cite{tsai},
\begin{equation}
R_{V/A;ij}\, =\, 12\pi^2\vert V_{ij}\vert^2 S_{EW}\, 
\int^{m_\tau^2}_0\, {\frac{ds}{m_\tau^2}} \,
\left[ w_T^{(0,0)}(y_\tau )
\rho_{V/A;ij}^{(0+1)}(s) - w_L^{(0,0)}(y_\tau )\rho_{V/A;ij}^{(0)}(s) \right]
\label{basictaudecay}\end{equation}
where $y_\tau =s/m_\tau^2$, $w_T^{(0,0)}(y)=(1-y)^2(1+2y)$,
$w_L^{(0,0)}(y)=2y(1-y)^2$, $V_{ij}$ is the flavor $ij$ CKM matrix element, 
and $S_{EW}$ is a short-distance electroweak correction.
$(0+1)$ denotes the sum of $J=0$ and $1$ contributions.

Using the general finite energy sum rule (FESR) relation,
\begin{equation}
\int_{th}^{s_0}ds\, w(s)\, \rho (s)\, =\, {\frac{-1}{2\pi i}}\,
\oint_{\vert s\vert =s_0}ds\, w(s)\, \Pi (s)\ ,
\label{basicfesr}\end{equation}
valid for any correlator without kinematic singularities and any
analytic weight, $w(s)$, the RHS in Eq.~(\ref{basictaudecay}) can be 
replaced by an integral around $\vert s\vert = s_0=m_\tau^2$ 
involving the correspondingly weighted linear combination of 
$\Pi_{V/A;ij}^{(0,1)}(s)$, 
the $J=0,1$ scalar parts of the relevant current-current 
correlator. For large enough $s_0$, this integral can be evaluated using the
OPE~\cite{bnp}. Similar FESRs can be constructed for any $s_0<m_\tau^2$, 
either of $\Pi^{(0+1)}_{V/A;ij}(s)$ or $s\Pi^{(0)}_{V/A;ij}(s)$ and general 
analytic $w(s)$. The corresponding spectral integrals 
will be denoted generically by $R^w_{ij}(s_0)$ in what follows. We will
also use the term ``longitudinal'' to refer to the 
$J=0$ contribution in any $J=0+1/J=0$ decomposition.

Flavor-breaking differences of the $ij=ud,us$ spectral data,
\begin{equation}
\delta R^w(s_0)\, =\, \left[{\frac{R^w_{ud}(s_0)}{\vert V_{ud}\vert^2}}\right]
\, -\, \left[{\frac{R^w_{us}(s_0)}{\vert V_{us}\vert^2}}\right]\ ,
\label{tauvusbasicidea}\end{equation}
have OPE representations, $\delta R^w_{OPE}(s_0)$, beginning 
with a $D=2$ contribution $\propto m_s^2$. Taking $\vert V_{ud}\vert$ and 
any OPE parameters from other sources, Eq.~(\ref{tauvusbasicidea}) 
provides a determination of $\vert V_{us}\vert$ for any $w(s)$,
and any $s_0<m_\tau^2$ sufficiently large that (i) convergence of the 
integrated series of known OPE terms is good and (ii) contributions
from insufficiently well-known higher $D$ OPE contributions are small.
Explicitly,
\begin{equation}
\vert V_{us}\vert \, =\, \sqrt{{\frac{R^w_{us}(s_0)}
{{\frac{R^w_{ud}(s_0)}{\vert V_{ud}\vert^2}}
\, -\, \delta R^w_{OPE}(s_0)}}}\ .
\label{tauvussolution}\end{equation}
For the weights used in the literature, at scales $s_0\sim 2-3\ {\rm GeV}^2$, 
$\delta R^w_{OPE}(s_0)$ is {\it much} smaller 
than $R^w_{ud,us}(s_0)$ (at the few-to-several-$\%$ level). 
The fractional uncertainty on $\vert V_{us}\vert$ 
($\simeq \Delta \left(\delta R^w_{OPE}(s_0)\right)/2\, R^w_{ud}(s_0)$)
produced by an uncertainty 
$\Delta \left(\delta R^w_{OPE}(s_0)\right)$ in $\delta R^w_{OPE}(s_0)$ 
is thus {\it much} smaller than that on $\delta R^w_{OPE}(s_0)$ itself,
allowing modest precision in the determination of $\delta R^w_{OPE}(s_0)$
to be turned into a high precision determination of 
$\vert V_{us}\vert$~\cite{gamizetal}.
Two important theoretical complications, however, must be dealt with.

The first complication is the very bad convergence of the integrated 
longitudinal $D=2$ OPE series~\cite{convproblem} 
and strong spectral-positivity-constraint violations produced 
by all truncation schemes employed in the literature
for this badly converging series~\cite{longposviol}. These problems 
necessitate removing longitudinal contributions from the experimental 
spectral distributions so that one can work exclusively with
sum rules based on the better-behaved $J=0+1$ OPE combination. 
Fortunately, for a combination of chiral and kinematic reasons, 
the spectral subtraction is strongly dominated by the
well-known $\pi$ and $K$ pole contributions. The remaining
non-pole, or ``continuum'', contributions are doubly chirally suppressed,
by factors of $O\left( [m_d\pm m_u]^2\right)$ (respectively, 
$O\left( [m_s\pm m_u]^2\right)$) for the flavor $ud$ 
(respectively $us$) V/A channels. The $ij=ud$ non-pole longitudinal
subtraction is thus numerically negligible. For $ij=us$, 
the continuum part of the longitudinal subtraction, though
small, is not entirely negligible, but can be determined
using input from sum rule and dispersive studies of the $us$
scalar and pseudoscalar channels~\cite{jop,mksps}.
The normalization of these results is strongly constrained by independent
information on $m_s$ from the lattice and hence under good control 
theoretically. The resulting $us$ continuum longitudinal subtraction is, as 
expected, numerically small.

The second complication is the less than optimal convergence
of the $J=0+1$ $D=2$ OPE series. Defining $ \Delta\Pi (s)\, 
\equiv\, \left[ \Pi_{V+A;ud}^{(0+1)}(s)\, -\,
\Pi_{V+A;us}^{(0+1)}(s)\right]$, one has explicitly, 
with $\alpha_s(Q^2)$ and $m_s(Q^2)$ the running coupling and strange quark 
mass in the $\overline{MS}$ scheme, and 
$\bar{a}=\alpha_s(Q^2)/\pi$~\cite{bck05},
\begin{eqnarray}
\left[\Delta\Pi (Q^2)\right]^{OPE}_{D=2}\, &=&\, {\frac{3}{2\pi^2}}\,
{\frac{m_s(Q^2)}{Q^2}} \left[ 1\, +\, {\frac{7}{3}} \bar{a}\, +\, 
19.93 \bar{a}^2 \, +\, 208.75 \bar{a}^3
\right. \nonumber\\ 
&&\left. \qquad\qquad\qquad
\ \ \ +\, (2378\pm 200)\bar{a}^4\, +\, \cdots \right]\ ,
\label{d2form}\end{eqnarray}
where the $\bar{a}^4$ coefficient is an estimate obtained using methods
which previously provided a rather accurate estimate of the $\bar{a}^3$ 
coefficient in advance of the actual calculation of its value~\cite{bck05}.
Since independent high-scale determinations of $\alpha_s(M_Z)$ correspond to
$\bar{a}(m_\tau^2)\simeq 0.10$, 
convergence at the spacelike point on
$\vert s\vert =s_0<m_\tau^2$ is marginal at best. While the decrease of
$\vert \alpha_s(Q^2)\vert$ on moving away from the spacelike point
along the contour {\it does} allow the convergence of the integrated
series to be improved through judicious choices of the
sum rule weight $w(s)$, for weights not chosen specifically 
with this purpose in mind very slow $D=2$ convergence is 
to be expected. Such slow convergence is, for
example, seen explicitly in the case of the so-called $(k,m)$ spectral 
weights~\cite{bck05}. Because of the growth of $\alpha_s$ with
decreasing scale, premature truncation of such a slowly converging series 
will produce $\vert V_{us}\vert$ values with an unphysical $s_0$-dependence. 
In view of the intrinsically slow convergence of the underlying
$D=2$ correlator series, a study of the $s_0$-dependence of the output 
$\vert V_{us}\vert$ (and demonstration of the existence of an
$s_0$-stability window) is thus crucial to establishing the reliability of
any hadronic-$\tau$-decay-based FESR extraction of $\vert V_{us}\vert$. 
Analyses in which such a study has not been performed thus typically
contain additional theoretical systematic uncertainties, which may
be significantly larger than the theoretical errors explicitly studied. 

\begin{table}[th]
\caption{$\tau^-\rightarrow X^-_{us}\nu_\tau$ branching fractions for strange
hadronic states $X^-_{us}$. Column 2 contains the PDG2006 world averages,
column 3 the 2007 Babar and/or Belle results, and column 4 the
resulting new world averages, with PDG-style $S$ factors 
included where needed.}
\vskip .05in
{\begin{tabular}{@{}llll@{}}
\toprule
$X^-_{us}$&${\cal{B}}_{WA,2006}\, (\%)$&${\cal{B}}_{2007}\, (\%)$
&${\cal{B}}_{WA,2007}\, (\%)$\\
\hline
$K^-$ [$\tau$ decay]&$0.691\pm 0.023$&&$0.691\pm 0.023$  \\
\ \ \ \ \ \ ([$K_{\mu 2}$])&($0.715\pm 0.003$)&&($0.715\pm 0.003$) \\
$K^-\pi^0$&$0.454\pm 0.030$&$0.416\pm 0.018$~\cite{babar07}&$0.426\pm 0.016$ \\
$\bar{K}^0\pi^-$&$0.878\pm 0.038$&$0.808\pm 0.026$~\cite{belle07}
&$0.831\pm 0.028$ ($S=1.3$)\\
$K^-\pi^0\pi^0$&$0.058\pm 0.024$&&$0.058\pm 0.024$\\
$\bar{K}^0\pi^0\pi^-$&$0.360\pm 0.040$&&$0.360\pm 0.040$ \\
$K^-\pi^-\pi^+$&$0.330\pm 0.050$&$0.273\pm 0.009$~\cite{babar07}
&$0.280\pm 0.016$ ($S=1.9$) \\
$K^-\eta$&$0.027\pm 0.006$&$0.0162\pm 0.0010$~\cite{belle07}
&$0.016\pm 0.002$ ($S=1.8$)\\
$(\bar{K}3\pi )^-$ (est'd)&$0.074\pm 0.030$&&$0.074\pm 0.030$ \\
$K_1(1270)\rightarrow K^-\omega$&$0.067\pm 0.021$&&$0.067\pm 0.021$ \\
$(\bar{K}4\pi )^-$ (est'd)&$0.011\pm 0.007$&   &$0.011\pm 0.007$ \\
$K^{*-}\eta$&$0.029\pm 0.009$&$0.0113\pm 0.0020$~\cite{belle07}
&$0.012\pm 0.004$ ($S=2.0$) \\
$K^-\phi$&&$0.00370\pm 0.00025$~\cite{babar07,belle07}
&$0.00370\pm 0.00025$ \\
\hline
TOTAL&$2.979\pm 0.086$&&$2.830\pm 0.074$ \\
&($3.003\pm 0.083$)&& ($2.854\pm 0.071$) \\
\hline
\end{tabular} \label{table1}}
\end{table}

\section{The updated determination of $\vert V_{us}\vert$}
Our results for $\vert V_{us}\vert$ are based on the ALEPH
$ud$~\cite{alephud05} and $us$~\cite{alephus99} spectral data,
for which information on the relevant covariance matrices is
publicly available. Since no
V/A separation has been performed for the $us$ data, we work
with the V+A combination (the $ud$ spectral integral errors are
also reduced by working with the V+A combination). 
The older $us$ data~\cite{alephus99} has been modified to 
incorporate the recent (2007) BaBar~\cite{babar07} and Belle~\cite{belle07}
updates of a number of the strange branching fractions, ${\cal{B}}_{X_{us}}$. 
The latter measurements, together with the previous and
resulting world averages, are shown in Table~\ref{table1}. 
Since the re-measurement of the strange spectral 
distribution is not yet complete, we follow 
Ref.~\cite{alephrescale} and obtain a partially updated
inclusive sum by rescaling the 1999 ALEPH distribution~\cite{alephus99},
mode-by-mode, by the ratio of new to old ${\cal{B}}$ values. 

\begin{figure}[t]
\rotatebox{270}{\mbox{
\centerline{\psfig{file=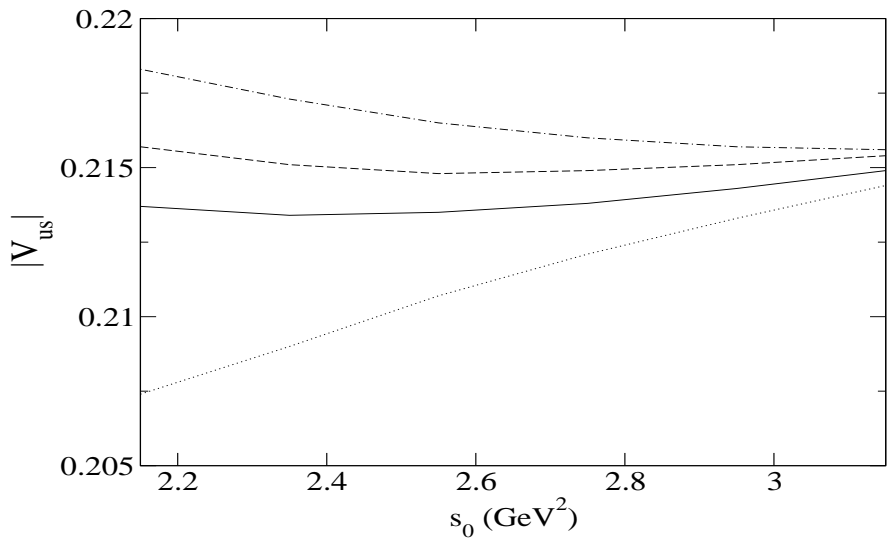,height=13.5cm,width=8.5cm}
}}}
\caption{$\vert V_{us}\vert$ versus $s_0$ for, from top to bottom,
$w_{20}$, $\hat{w}_{10}$, $w_{10}$ and $w_T^{(0,0)}$. \label{figure1}}
\end{figure}

With the above $ud$ and $us$ V+A spectral distributions as input,
$\vert V_{us}\vert$ is determined, as a function of the upper integration
limit, $s_0$, in Eq.~(\ref{basicfesr}), using $J=0+1$ FESRs based on 
(i) the three improved-convergence, non-spectral weights, 
$w_{10}$, $\hat{w}_{10}$ and $w_{20}$, introduced in Ref.~\cite{km00}, 
and (ii) the kinematic weight $w_T^{(0,0)}$ (expected to display
problematic integrated OPE convergence behavior and
hence $s_0$-instability for $\vert V_{us}\vert$~\cite{kmcw,bck05}). 
Standard OPE input, the correlator form of the $D=2$ contributions,
and $K_{\mu 2}$ input for the $K$ pole contribution to the
$us$ spectral integrals are employed. 
More details of the analysis will be reported elsewhere~\cite{mwbnr08}. 
The resulting $s_0$ dependence of $\vert V_{us}\vert$ 
is shown in Fig.~\ref{figure1}.

The figure shows reasonable $s_0$-stability for $w_{10}$, $\hat{w}_{10}$
and $w_{20}$ in the region above $s_0\sim 2.5\ {\rm GeV}^2$, 
but significant $s_0$-instability for $w_T^{(0,0)}$.
The $s_0=m_\tau^2$ $\vert V_{us}\vert$ values, 
\begin{eqnarray}
&&0.2156\pm 0.0028_{exp}\pm 0.0022_{th}\ \ {\rm for}\ w_{20}\nonumber\\
&&0.2154\pm 0.0032_{exp}\pm 0.0015_{th}\ \ {\rm for}\ \hat{w}_{10}\nonumber\\
&&0.2149\pm 0.0033_{exp}\pm 0.0010_{th}\ \ {\rm for}\ w_{10}\nonumber\\
&&0.2144\pm 0.0030_{exp}\pm 0.0017_{th}\ \ {\rm for}\ w_T^{(0,0)}\ ,
\label{vusresults}\end{eqnarray}
are in good agreement, but lie $\sim 3\sigma$ below the
expectations of $3$-family unitarity~\cite{newht07}. The
theoretical errors have been estimated using the scheme
discussed in Refs.~\cite{kmcw}, which is, by design, more conservative
than that employed in previous analyses. In spite of this,
the $s_0$-instability of the results from the $w_T^{(0,0)}$ 
analysis suggests that the theoretical error in this case may
still be insufficiently conservative. The $\sim~0.0050$
downward shifts in $\vert V_{us}\vert$, compared to the 2006 versions of these 
analyses~\cite{gamizetal,kmcw}, result dominantly from the 
$-0.075\%$ ($-0.050\%$) shifts in ${\cal{B}}_{K\pi}$ 
(${\cal{B}}_{K^-\pi^+\pi^-}$), which represent, respectively,
$2.6\%$ ($1.8\%$) decreases in the total strange branching
fraction and hence, from Eq.~(\ref{tauvussolution}), 
$\sim 1.3\%$ ($0.9\%$) downward shifts in $\vert V_{us}\vert$.

\begin{figure}[t]
\centerline{\psfig{file=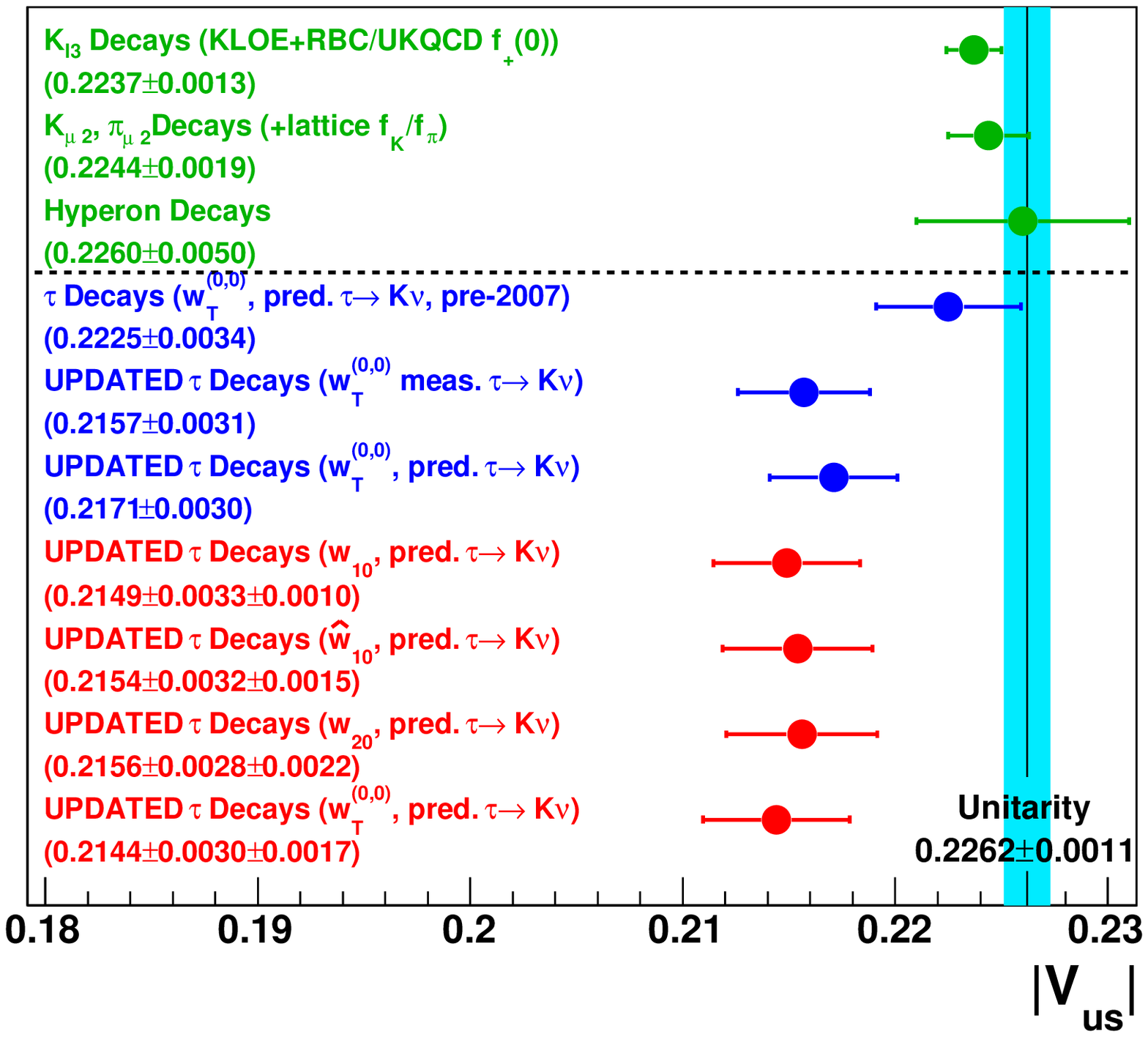,
height=16.5cm,width=14.5cm}}
\vspace*{8pt}
\caption{$\vert V_{us}\vert$ from various sources \label{figure2}}
\end{figure}

In Figure~\ref{figure2} we compare the results above for $\vert V_{us}\vert$
to those of a number of other determinations, from various sources. The 
result from $K_{\ell 3}$ decay is based on the most recent KLOE $K$ decay 
data~\cite{kloek08}, together with the RBC/UKQCD lattice determination of 
$f_+(0)$~\cite{rbcukqcdfplus07}, that for Marciano's
method (which employs the ratio of
$K_{\mu 2}$ to $\pi_{\mu 2}$ decay widths in combination with 
lattice input for $f_K/f_\pi$~\cite{marcianovus})
on the FlaviaNet Kaon Working Group assessment of the
experimental ratio~\cite{fnwgk08} and Juttner's Lattice 2007 
plenary assessment of the status of lattice determinations of 
$f_K/f_\pi$~\cite{juttner07}, while that from hyperon decays,
included for completeness, is Jamin's assessment as of Electroweak 
Moriond 2007 (which is $0.0010$ higher, and with larger errors, 
than the result quoted in Ref.~\cite{csw03}). The first of the $\tau$ decay
results represents a range of assessments based on the $w_T^{(0,0)}$
FESR and pre-2007 $us$ data. The next two $w_T^{(0,0)}$ FESR
assessments are from Ref.~\cite{banerjee07} and incorporate
most, but not all, of the new 2007 $us$ data. The
sizeable difference produced by switching from 
$K_{\mu 2}$ input for the $K$ pole contribution (the
``pred. $\tau\rightarrow K\nu$'' result) to the central measured
value (the ``meas. $\tau\rightarrow K\nu$'' result) is a reflection of
the high degree of cancellation between the $ud$ and $us$
spectral integrals~\cite{pichalternatefootnote}.
Note that the $\tau$ decay results just mentioned, as well as those
of Refs.~\cite{pichetal}, employ the 4-loop-truncated Adler 
function evaluation of the $D=2$ OPE integral contribution. 
The remaining $\tau$ decay results shown in the figure
are the ones obtained by us above, which, in contrast, employ 
the 4-loop-truncated correlator evaluation of the $D=2$ contribution.
The non-trivial difference between the results
obtained using the two different methods (which are formally
equivalent, up to contributions of $O(\alpha_s^4)$ and higher)
provides further evidence for the slow convergence of the
integrated $D=2$ series for the $w_T^{(0,0)}$ weight, 
seen already in Fig.~\ref{figure1}.

While the results above are compatible with
small beyond-the-SM contributions, such a conclusion is obviously 
premature until BaBar and Belle have 
cross-checked each other's recent results and also
filled in the missing entries in column 3 of Table~\ref{table1}
(especially for the sizable $\bar{K}^0\pi^0\pi^-$ and previously 
estimated, but unmeasured, $\bar{K}3\pi$ and $\bar{K}4\pi$ modes). 
Results for higher multiplicity modes, with branching fractions
down to the few $\times 10^{-5}$ level, are also needed for
a determination at the desired level of accuracy. A finalized
version of the results will require completion of the re-measurement 
of the full $us$ spectral distribution, work on which is
in progress. Note that improved precision on the measurement
of the ratio of branching fractions of the single-particle $\pi$ and $K$ 
modes is also of interest for the determination of $\vert V_{us}\vert$
since it would allow an independent detemination in the $\tau$ sector
analogous to that based on 
$\Gamma [K_{\mu 2}]/\Gamma [\pi_{\mu 2}]$~\cite{marcianovus}. 
Experimental input on the 
chirally-suppressed continuum $us$ $J=0$ spectral contributions,
needed for performing the longitudinal subtraction, appears potentially
experimentally feasible, and would also be most welcome. 

Finally, we note that a re-measurement of the $ud$ distribution is also 
of relevance, particularly in light of the known discrepancy between
the $\pi\pi$ and $4\pi$ distributions measured in $\tau$ decay and 
those implied by electroproduction cross-sections (via CVC and
known isospin-breaking corrections). Were the
electroproduction expectations to be correct, the values of 
$\vert V_{us}\vert$ obtained from the above analyses would be
increased by $\sim 0.0018$. It is worth noting, in this regard,
that two recent analyses of non-strange hadronic $\tau$ decay 
data~\cite{jb08,my08}, one  
designed specifically to incorporate an improved treatment of $D>4$ OPE 
contributions which were problematic for earlier determinations, 
yield values of $\alpha_s(M_Z)$ in excellent agreement with the 
results, $0.1190(26)$ and $0.1191(27)$, 
obtained from updates to the global fit to electroweak observables 
at the Z scale (quoted in Refs.~\cite{bck08,davier08}, respectively).
Two recent updates~\cite{hpqcd08,mlms} of the earlier lattice
determination~\cite{hpqcdukqcd05}, employing slightly different 
implementations of the original approach, also obtain values ($0.1183(7)$ and 
$0.1192(11)$, respectively) which are higher than the original result 
and now in excellent agreement with the results of the $\tau$ extraction and 
global electroweak fit.
Since electroproduction cross-sections imply 
a significantly lower value, $\alpha_s(M_Z)\sim 0.1140-0.1150$~\cite{km05}, 
we consider it unlikely that the $ud$ $\tau$ data is significantly in
error, but nonetheless a re-measurement from the B factory experiments
should be pursued, given the potential impact on both SM expectations
for $(g-2)_\mu$ and on the value of $\vert V_{us}\vert$ obtained
in the hadronic $\tau$ decay analyses.

\section{Acknowledgments}

KM and JMR acknowledge the ongoing support of the Natural Sciences and
Engineering Research Council of Canada. KM also acknowledges the
hospitality of the Theory Group at IHEP, Beijing and the CSSM
at the University of Adelaide.


\begin{thebibliography}{0}
\bibitem{pichalternatefootnote}The $K_{\mu 2}$-based result shown
in the Figure, $0.2171\pm 0.0030$, differs slightly from that
obtained in Refs.~\cite{pichetal} ($0.2165\pm 0.0026$) as a result 
of the use of slightly different OPE input.
\end{thebibliography}
\end{document}